# Role of Zealots on the Adaptive Voter Model

**Ka Wai Cheung[1], Chung Him Liu[1] and Kwok Yip Szeto[1]**

Department of Physics, Hong Kong University of Science and Technology
phszeto@ust.hk

**Abstract.** The voter model has been extensive studied as an opinion dynamic model, and the role of the zealots has only been discussed recently. We introduce the adaptive voter model with zealots and show that the final distribution of the magnetism can be separated into two regions depending on the number of zealots as well as the probability of forming link. When the fraction of zealots is dominated in the population, the probability distribution of magnetism follows a Gaussian-like distribution and the relaxation time is population-size independent. When the population is dominated by the susceptible agents, the relaxation time is proportional to the exponential of the population size. We have found the analytical solution of the relaxation time in the limiting cases and explained the difference of the relaxation time in these two regions based on the approximation method.

## 1. Introduction

Opinion dynamic is a study of how the interaction among individuals forms a collective behaviour, and the voter model is the most basic and perhaps the completely solved example of this kind of behaviour [1]. In the voter model, each individual, generally known as voter, can switch between two opinion states based on his neighbours' opinions. The perfect partisan voters, or known as the 'zealots' or 'pinning controller', may never changes his opinion and guide the states of the other voters to a desired state [2]. In the social network, the zealot usually corresponds to the leader in the community, the mass media or the regular voters in the election. And in the voter model, the existence of the zealot favour of opposing opinions like competing parties, which will introduce a non-consensus state in equilibrium. The idea of the zealot on the voter model is first proposed by Mobilia [3]. They have studied the effect of single zealot on the hypercubic lattice, and showed that single zealot cannot influence all individuals when the dimension of the lattice is larger than two. The effect of multiple zealots on the homogenous and static graph is later analysed by Mobilia *et.al.* [4] and Yildiz *et.al.* [5], and the role of zealots on the variations of voter model is studied by Düring *et.al.* [6], Khalil *et.al.* [7] and Acemoğlu *et.al.* [8].

In this paper we want to study the effect of zealots on a coevolving-network [9,10], where the links and the opinions of the voters are updating simultaneously, the key question is that how does the role of zealot affect the equilibrium state as well as the convergence time. In this paper we will first introduce a mathematical solvable model, and focus on the symmetric case when the number of opposite zealots is equal.

## 2. Model

In the adaptive voter model, each individual has an opinion which takes two values $s_i = \pm 1$. At each time step a randomly chosen voter is undergoing two stages: link update follows by an opinion update. For the link update stage, the voter removes all links from his neighbours and reforms the links with *all* other nodes with two fixed probabilities $q_0$ and $q_1$, where $q_0$ is the probability such that two agents holding the same opinion will establish a link, and $q_1$ is the probability such that two agents holding the opposite opinion will form a link. Because $q_0$ and $q_1$ corresponds two separated events, there is no necessary constraint between these two control parameters. However, a simplification can be made by assuming $q = q_0 = 1 - q_1$. When $q > 0.5$, the voter tends to form links with agents holding the same opinion, which demonstrates the Homophily behaviour in the society; when $q < 0.5$, the voter tends to form links with agents holding the opposite opinion, which may demonstrate the Heterophily behaviour. At the opinion update stage, we follow the node-updated rule stated in [11]. If the voter is normal (susceptible), he will randomly pick one of his neighbours and follow his opinion; if the voter is a zealot, his opinion is kept unchanged. We should point out that there are two major differences compared to the original adaptive voter model [9], the first difference is that we introduce the role of zealot in the network and will later shows its importance in the opinion dynamic; the second difference is that original adaptive voter model uses the majority-update rule in the opinion-update stage, and here we use the node-update rule, which leads to a mathematical solvable model.

We focus on the time evolution of the probability distribution of the average opinion $m(t)$ of the *entire population* $N$, which is can be written as $m(t) = \sum s_i(t)/N$. Let $n$ be the fraction of normal agents holding the positive opinion, $z_+(z_-)$ be the fraction of the zealots holding the positive (negative) opinion. The average opinion is then can be obtained as

$$m(t) = 2n(t) - s + z_+ - z_-  \qquad (1)$$

where $s = 1 - z_+ - z_-$. The time evolution of the distribution $n(t)$ can be approximated by the Fokker-Planck equation

$$\frac{\partial p(n,t)}{\partial t} = -\frac{\partial}{\partial n}\big(\alpha(n)p(n,t)\big) + \frac{1}{2}\frac{\partial}{\partial n}\big(\beta(n)p(n,t)\big) \tag{2}$$

where $a(n) = W(n + 1/N|n) - W(n - 1/N|n)$ and $\beta = (W(n + 1/N|n) + W(n - 1/N|n))/N$. $W(n + 1/N|n)$ and $W(n - 1/N|n)$ is the transition probability density from $n$ to $n + 1/N$ and $n - 1/N$, which corresponds to the birth rate and death rate of $n$ separately. In the adaptive voter model with node-update rule, the transition probability is equal to

$$W(n + 1/N|n)\Delta t = \frac{q_1(s - n)(n + z_+)}{q_1(n + z_+) + q_0(s - n + z_-)}$$

$$W(n - 1/N|n)\Delta t = \frac{q_1 n(s - n + z_-)}{q_1(s - n + z_-) + q_0(n + z_+)}$$

(3)

and $\Delta t$ is chosen to be $1/N$ in the Monte-Carlo simulation. $W(n + 1/N|n)$ is the conditional probability such that the positive spin is increment by 1, and it can be decomposed into a product of two probabilities: the probability of choosing a negative spin and the probability of choosing one of positive spin from that negative spin's neighbourhood. Same factorization can be applied to $W(n - 1/N|n)$. From the statistical perspective, $q_0$ ($q_1$) controls *how likely* a pair of same (opposite) spins is picked from a randomly distributed environment. When $q_0$ is higher, the probability of picking a pair of same spin is higher. In this sense, $q_0$ and $q_1$ can be viewed as weights in the sampling. When $q_0 = q_1$, the probability of picking a pair of same or opposite spins is same, the effect of $q_0$ and $q_1$ is cancelled out and left the transition probability constant.

## 3. Result

We focus on the simplified case such that $q = q_0 = 1 - q_1$. In this sense there are only three control parameters $\{q, z_+, z_-\}$, it will reduce to the classical voter model with zealot if and only if $q = 0.5$. When $z_+ = z_-$, the average opinion will be zero in the long-term limit. When $z_+ \neq z_-$, the majority's opinion is depending on the difference between $z_+$ and $z_-$.

### *3.1 The equilibrium distribution*

We consider the symmetric case where $z = z_+ = z_-$. The equilibrium distribution of $p(n)$ can be obtained by setting the time derivate of $p(n,t)$ to be zero. Based on the numerical calculation, it is found that the distribution of the $m$ heavily depends on the number of zealots, and it can be generally classified into two cases: a unimodal distribution or bimodal distribution as shown in Fig. 1.

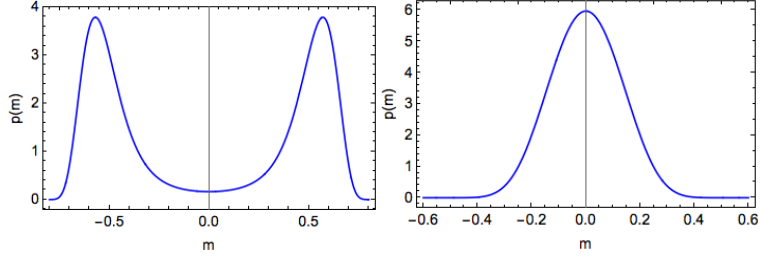

**Fig. 1.** The equilibrium distribution of the average opinion. (Left) $z = 0.1$. (Right) $z = 0.2$. Other parameters: $N = 200, q = 0.7$.

When the number of zealots is large, the distribution of the average opinion shows a single peak at $m = 0$ , which results in a unimodal distribution. When the number of zealots is small, the distribution of the average opinion has two peaks centred around the real roots of the $\alpha(n)$ , which corresponds to a bimodal distribution. The transition from the bimodal distribution to the unimodal distribution can be obtained by considering the slope of the equilibrium distribution. (See Appendix A) In the thermodynamic limit case where $N \to \infty$, we can obtain the boundary of the transition, which is

$$z_c = \frac{1}{2} - \frac{1}{4q} \qquad (4)$$

Under the assumption such that the population size is large, when $z > z_c$, the opinion in the entire population is well-mixed; when $z < z_c$, the society is dominated by a single opinion. Since $q \in (0,1)$, when $q < 0.5, z_c > 1$ which means that the distribution is always unimodal. The reason of this phase change can be explained based on the realization from the Monte-Carlo simulation. When $z$ is large, at the link-update stage, a randomly chosen voter will establish links with zealots with high probability. Because there exist zealots with opposite opinion, this voter still has high probability to change his opinion. However, when $z$ is small, all neighbours of this chosen voter have the same opinion, in this case the voter is unlikely changing his opinion in this Monte-Carlo step.

### *3.2 The dynamic evolution of the distribution*

The time scale approaches to the equilibrium or consensus is also an important topic in opinion formation. Consider an initial delta distribution where $< m >= m_0$, we can measure the time scale reaching the equilibrium state based on the numerical calculation. For simplicity, we can also measure the time scale based on the first moment of the average opinion distribution, which is defined as $< m >= \int mp(m, t) \, dm$. Result shows that when $z$ is large, the first moment quickly decays to a zero value; when $z$ is small, the first moment will be diverge to the metastable point shown in Fig. 2 and then decay to zero value in the long-time scale.

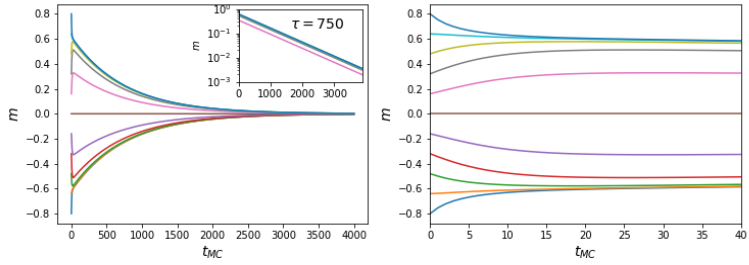

**Fig. 2.** The time evolution of the first moment of average opinion distribution for different initial delta distributions (shown as different colour). Fig. 2(a) shows long-time behaviour of the first moment; Fig. 2(b) shows the short-time behaviour.

The time evolution of the first moment can be obtained under the assumption such that the population size is infinitely large. In this case the motion of the distribution is governed by a first-order partial differential equation where the diffusion term involving $\beta(n)$ is ignored. Consider a linear approximation $\tilde{\alpha}(n)$ of the drift term $\alpha(n)$ at $n = s/2$, (See Appendix B) which is

$$\tilde{\alpha}(n) = [-8zq(1 - q) - 2(1 - q)(1 - 2q)](n - \frac{s}{2}) \qquad (5)$$

And also consider an initial delta distribution $p(n, 0) = \delta(n - \frac{s}{2} + \epsilon)$, where $|\epsilon| > 0$. Using the method of characteristics, we obtain

$$m = m_0 \exp(ct) , c = -8zq(1 - q) - 2(1 - q)(1 - 2q) \qquad (6)$$

When $c < 0$, the average opinion converges to 0 directly; when $c > 0$, the average opinion diverges from $m_0$ with speed $cm$. Since in the long-term limit $m(\infty) = 0$, there must exist a turning point $t^*$ such that $\dot{m}(t^*) = 0$, which corresponds to the metastable state. Solving $c = 0$, we obtain the equation of phase transition stated in equation (4). Based on these two facts, we can get the mapping shown in table 1.

**Table 1.** The equilibrium distribution and dynamic evolution in two regions

| Region | Equilibrium distribution | Dynamic evolution |
|---|---|---|
| $0 < z < z_c$ | Bimodal | Diverge to metastable state, and then converge to stable state. |
| $z_c < z < \frac{1}{2}$ or $q < \frac{1}{2}$ | Unimodal | Converge to stable state. |

The time to equilibrium in these two regions can be obtained by using the linear approximation in equation (5). When $z_c < z$, the relaxation time towards to the equilibrium can be approximated by equation (6), and the numerical result shows that the linear approximation fits the curve well when $z \to 1/2$, which is shown in Fig. 3(a). And we can see that the relaxation time is inversely proportional to the number of zealots and also it is population size independent. When $q \to 0$, the term involves $z$ stated in equation (6) vanishes. Thus, the role of zealot is not important compared to the case when $q \to 1$. When $z_c > z$, the divergence of the average opinion can also be approximated by equation (6), which is shown in Fig. 3(b). When the population size is large, the distribution will be centred at the real roots $n^*$ of $\alpha(n)$, using equation (1), we obtain two attractors for the probability distribution

$$m^* = \pm \sqrt{\frac{2q(1-2z)-1}{2q-1}} \qquad (7)$$

When $z$ increases, the peak of the metastable distribution will be shifted towards to 0. The time scale from the metastable state to the stable state also follows a decay function, which can be written as $m(t) = m(t_0)\exp(-t/\tau)$, $\tau = \exp(\lambda(q,z)N)$. $\lambda(q,z)$ is a nontrivial nonlinear function in terms $q, z$, the value can be obtained from the slope in Fig. 4(b). When the population size is infinite, the distribution will be stayed at the metastable state. When the population size is finite, an increment of size of population leads to a higher relaxation time. The finite size effect for both small $z$ and large $z$ is shown in Fig. 4.

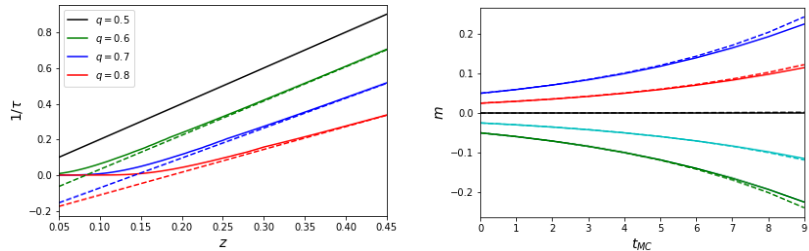

**Fig. 3.** Two figures show the linear approximation value (dashed line) and the numerical result from the differential equation (solid line). Fig 3(a) shows the inverse relaxation time versus the number of zealots with $N = 100$; Fig 3(b) shows the time evolution of the first moment of average opinion when $z = 0.05, q = 0.75, N = 2000$.

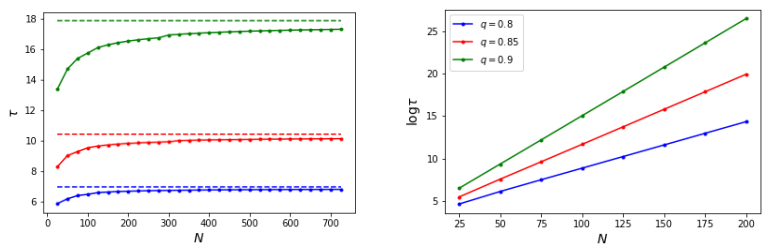

**Fig. 4.** The population size versus the relaxation time for different parameter $q$. Fig 4(a) shows the relaxation time when $z = 0.3$. The solid line is the numerical result obtain from equation (6), the dashed line is the theoretical limit. The relaxation time is lower than the theoretical limit when the population size is finite. Fig. 4(b) shows the logarithmic of relaxation time from the numerical result when $z = 0.1$, it shows a linear relation between logarithmic of relaxation time and the population size.

## 4. Discussion

In this paper we have analysed the effect of zealots on the adaptive voter model, especially in the symmetric case when the number of opposite zealots is the same. Because the coefficient $\alpha, \beta$ is nonlinear in terms $n$, it is generally difficult to obtain the analytical solution directly. Based on the approximation method, we have shown that the relaxation time obtained from the approximation fit the numerical result well when $z > z_c$, especially it will be the exactly solution when $z \rightarrow 0.5$. However, the approximation no longer holds when $z < z_c$ because of the existence of the metastable state. When $z < z_c$, the initial distribution of the average opinion will be attracted by one of attracters stated in equation (7) because the effect of drift term involves $\alpha$ is dominated in the dynamic evolution. Once the distribution is in the metastable state, the effect the drift term is much smaller. Because of the effects from the both drift term and diffusion term, another peak centred at the second attractor will appear and growth until the distribution is becoming symmetric. In the real-life application, because of the large population size, the metastable state will stay in a much longer time, thus the metastable state will be relatively important compared to the equilibrium state.

When the number of zealots is not same, there are three control parameters $\{z_+, z_-, q\}$. Rewrite it as $u = z_+ + z_-; v = (z_+ - z_-)/u$, where $u$ controls the total number of zealots with the range from 0 to 1, and $v$ controls the relative difference between two kinds of zealots, where it is ranged from -1 to 1. In this case the distribution of the average opinion in the equilibrium state is no longer symmetric. Consider the equilibrium solution of the Fokker-Planck equation, one can show that under the condition such that $v$ is fixed, when $u$ is higher, the first moment of the distribution will move towards to $\pm 1$ depending on the sign of $v$; when $u$ is lower, the first moment will move towards to $0$. And we have also found that the metastable exist if and only if both $z_+, z_-$ are smaller than a threshold. If either one of the $z_+, z_-$ is large, the distribution quickly converges to the stable state. It is because once the number of zealots is large, there exists a high probability such that each voter has one or more zealots as his neighbourhoods. In this case the opinion of each voter must be affected by the zealots in the relatively short time. Therefore, the metastable states will not exist if the number of zealots is large.

## 5. Appendix

[A]. The boundary of these two regions can be obtained by considering the equilibrium solution of equation (2), which satisfies the relation

$$\frac{p'(n)}{p(n)} = \frac{\alpha(n) - \beta'(n)}{\beta(n)} + const \approx \frac{\alpha(n)}{\beta(n)} + const \tag{8}$$

The second approximation holds when $N \to \infty$ since $\beta(n)$ is in order of $1/N$. Given an equilibrium distribution shown in Fig.1., we can empirically find that the $p'(n)/p(n)$ has no turning point when $z$ is large, and it has two turning points when $z$ is small. Due to the fact that the equilibrium distribution is symmetric, we can deduce that the derivative of $\alpha(n)/\beta(n)$ at $n = s/2$ is less than zero if $z$ is large than a threshold $z_c$, otherwise it is larger than zero. Based on this fact, solve $\alpha'(n)/\alpha(n) = \beta'(n)/\beta(n)$ we can get the boundary equation stated in equation (4).

[B]. The linear approximation $\tilde{\alpha}(n)$ of $\alpha(n)$ is exact when $z \to s/2$, and it is also nontrivially exact when $q \to 0,1$. Denote the difference between $\tilde{\alpha}(n)$ and $\alpha(n)$ as $g(n)$, we can obtain

$$g(n) = \frac{K(2n - s)^3}{\Gamma(\Gamma + 1)} \tag{9}$$

where $K = -q(2q^2 - 3q + 1)(1 - z - q(1 - 2z))$ , $\Gamma = (-n - q - z + 2qn + 2qz)$. When $q \to 0$ or $1$, $K \to 0$ and $\Gamma(\Gamma + 1) \neq 0$. Thus $g(n) \to 0$ if $q \to 0,1$. Therefore, the linear approximation and its following derivations will also be exact under these two extreme cases.